\documentclass[11pt, onecolumn]{IEEEtran}
\usepackage{lmodern}
\usepackage[letterpaper, left=1in, right=1in, bottom=1in, top=1in]{geometry}

\usepackage{amsmath}  
\usepackage{amssymb}  
\usepackage{mathrsfs} 
\usepackage{amsthm}
\usepackage{bbm}

\usepackage{comment}  

\usepackage{upref}
\usepackage{amsfonts}

\usepackage{verbatim}

\usepackage[numbers,sort&compress]{natbib}

\usepackage[dvipsnames,usenames]{color}

\usepackage{graphicx}
\usepackage{subfig}
\usepackage{tikz}

\usepackage{latexsym}
\usepackage{makecell}
\usepackage{enumitem}

\usepackage{pgfplots}
\usepgfplotslibrary{dateplot}
\pgfplotsset{compat=1.8}

\allowdisplaybreaks

\newcommand{\be}[1]{\begin{equation}\label{#1}}
	\newcommand{\ee}{\end{equation}}

\newcommand{\bc}{\begin{center}}
	\newcommand{\ec}{\end{center}}

\newcommand{\cA}{{\cal A}}

\newcommand{\cI}{{\cal I}}

\newcommand{\cS}{{\cal S}}

\let\originallesssim\lesssim
\let\originalgtrsim\gtrsim

\DeclareRobustCommand{\lesssim}{%
  \mathrel{\mathpalette\lowersim\originallesssim}%
}
\DeclareRobustCommand{\gtrsim}{%
  \mathrel{\mathpalette\lowersim\originalgtrsim}%
}

\newcommand{\al}{\alpha}

\newcommand{\ep}{\epsilon}

\newcommand{\Cref}[1]{Co\-rol\-la\-ry\,\ref{#1}}

\newtheorem{theorem}{Theorem}
\newtheorem{cor}{Corollary}
\newtheorem{lemma}{Lemma}

\newcommand{\poly}{\text{poly}}

\newif\ifdraft

\draftfalse 
\drafttrue 

\newcommand{\aln}[1]{\begin{align*}#1\end{align*}}
\renewcommand{\al}[1]{\begin{align}#1\end{align}}

\renewcommand{\thesection}{\arabic{section}}

\usepackage{titlesec}
\titleformat{\section}
  {\normalfont\bfseries}
  {\thesection}{1em}{}

\begin{document}
	\sloppy
	
	\usetikzlibrary{shapes}
	\newcommand{\alphaVal}{0.01}

	\title{\Large Substring Density Estimation from Traces}

\author{Kayvon Mazooji \quad \quad Ilan Shomorony}
	
\author{\IEEEauthorblockN{}
\and
\IEEEauthorblockN{Kayvon Mazooji and Ilan Shomorony}
\IEEEauthorblockA{ \\
	Department of Electrical and Computer Engineering\\
	University of Illinois, Urbana-Champaign\\
	mazooji2@illinois.edu,  ilans@illinois.edu}
\and
\IEEEauthorblockN{}
\thanks{The work of Kayvon Mazooji and Ilan Shomorony was supported in part by the National Science Foundation (NSF) under grants CCF-2007597 and CCF-2046991.}
}

\date{}

\maketitle

	

	
	\maketitle

\begin{abstract} 
In the trace reconstruction problem, one seeks to reconstruct a binary string $s$ from a collection of traces, each of which is obtained by passing $s$ through a deletion channel.
It is known that $\exp(\tilde O(n^{1/5}))$ traces suffice to reconstruct any length-$n$ string with high probability.
We consider a variant of the trace reconstruction problem where the goal is to recover a ``density map'' that indicates the locations of each length-$k$ substring throughout $s$.
We show that $\ep^{-2}\cdot \poly(n)$ traces suffice to recover the density map with error at most $\epsilon$.
As a result, when restricted to a set of source strings whose minimum ``density map distance'' is at least $1/\poly(n)$, the trace reconstruction problem can be solved with polynomially many traces.


\end{abstract}  

	
	\section{Introduction}

In the trace reconstruction problem, there is an unknown binary string $s \in \{0,1\}^n$, which we wish to reconstruct based on $T$ subsequences (or traces) of $s$.
Each trace is obtained by passing the {\it source string} $s$ through a deletion channel, which deletes each bit of $s$  independently with probability $p$.
The main question of interest is how many traces are needed in order to reconstruct $s$ correctly.

This problem was originally proposed by Batu et al.~\cite{batu2004reconstructing}, motivated by problems in sequence alignment, phylogeny, and computational biology~\cite{bhardwaj2020trace}.
Most of the work on the trace reconstruction problem has focused on characterizing the minimum number of traces needed for reconstructing the source string $s$ exactly.
The most common formulation of the problem, known as
\emph{worst-case trace reconstruction}~\cite{Servedio}, 
requires the reconstruction algorithm to recover $s \in \{0,1\}^n$ exactly with high probability as $n \to \infty$ for \emph{any} string $s \in \{0,1\}^n$.
While this problem has received considerable attention, there is still a significant gap between upper and lower bounds on the number of traces needed.
Currently, the best lower bound
is $\tilde{\Omega}(n^{3/2})$, while the best upper bound
is $\exp(\tilde{O}(n^{1/5}))$, both due to Chase~\cite{chase2,chase}.

The exponential gap between the best known lower and upper bounds has motivated the formulation of several variants of the trace reconstruction problem where tighter bounds can hopefully be obtained.
For example, in the \emph{average-case trace reconstruction} problem, $s$ is assumed to be drawn uniformly at random from all $\{0,1\}^n$ strings.
In this case, it is known that only $T = \exp(O(\log^{1/3}(n)))$ traces are sufficient~\cite{Peres2}.
An \emph{approximate trace reconstruction} problem, where a fraction of the recovered bits is allowed to be incorrect, has also been formulated~\cite{davies2021approximate}, and the problem of finding the maximum likelihood sequence $s$ from a small number of traces (possibly insufficient for exact reconstruction) has been recently studied~\cite{Diggavi}.
We can also consider a more modest goal than the reconstruction of the source sequence $s$ itself.
One example that is particularly relevant to our discussion is the reconstruction of the $k$-subword deck of $s$~\cite{holenstein2008trace,Servedio2}.

\begin{figure} 
	\centering
	\includegraphics[width=0.9\linewidth]{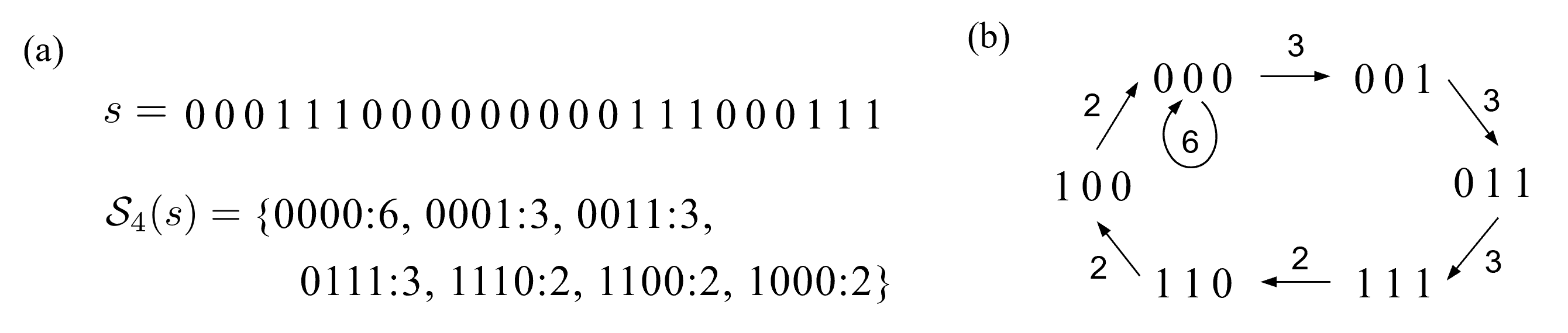}
	\caption{(a) Example of a source binary string $s$ and its $k$-subword deck (or $k$-spectrum) $\cS_k(s)$, for $k=4$.
	(b) Given $\cS_4(s)$ in (a), one can build a de Bruijn graph where the elements in $\cS_4(s)$ correspond to edges (with multiplicities) and the nodes correspond to $3$-mers.
	Notice that $s$ corresponds to an Eulerian path on the de Bruijn graph, but such a path is not unique; for example, $s' = 000111000111000000000111$ corresponds to another Eulerian path.
		\label{fig:debruijn}
	}
\end{figure}

The $k$-subword deck of a binary sequence $s \in \{0,1\}^n$ is the multiset of all length-$k$ substrings, i.e.,
$\{ s[i:i+k-1] : i =1,\dots,n-k+1 \}$.
Equivalently, the $k$-subword deck can be defined by the counts $N_{s,x}$ of the number of times $x$ appears in $s$ as a substring:
\al{
\cS_k(s) = [ N_{s,x} : x \in \{0,1\}^k ].
}
As shown in \cite{Servedio2}, for $k = O(\log n)$, the $k$-subword deck $\cS_k(s)$ can be recovered with $\poly(n)$ traces.
The $k$-subword deck of a sequence is an important object in  bioinformatics, with applications in error correction~\cite{liu2013musket,shomorony2016fundamental}, sequence assembly~\cite{EULER,marcovich2021reconstruction}, and genomic complexity analysis~\cite{chor2009genomic,TNF}.
In these contexts, the $k$-subword deck $\cS_k(s)$ is often referred to as the $k$-spectrum, and each length-$k$ substring is called a $k$-mer.
Intuitively, as long as $k$ is large enough, the $k$-subword deck can uniquely determine the source sequence $s$.
In fact, a classical result by Ukkonen~\cite{ukkonen1992approximate} provides a necessary and sufficient condition for $\cS_k(s)$ to uniquely determine $s$ based on the length of the ``interleaved repeats'' in $s$~\cite{BBT}.
In particular, if there are no repeats of length $k-1$ in $s$, one can reconstruct $s$ from $\cS_k(s)$ by simply merging $k$-mers with a prefix-suffix match of length $k-1$.
More generally, given $\cS_k(s)$, one can build the \emph{de Bruijn graph}, where nodes correspond to $(k-1)$-mers and edges correspond to $k$-mers, and $s$ is guaranteed to be an \emph{Eulerian path} in the graph~\cite{EULER,shomorony2016information}.
This is illustrated in Figure~\ref{fig:debruijn}.

While the $k$-subword deck is a natural intermediate goal towards the reconstruction of $s$ (and can be recovered with only $\poly(n)$ traces),
it does not capture all the information present in the traces.
For example, the $k$-subword deck $\cS_k(s)$ in Figure~\ref{fig:debruijn} also admits the reconstruction $s' = 000111000111000000000111$, even though $s'$ should be easy to distinguish from $s$ based on traces (by estimating the length of the second and third runs of zeros).
Motivated by this shortcoming of the $k$-subword deck, we propose the idea of a $k$-mer density map, as a kind of localized $k$-subword deck where, in addition to knowing the number of times a given $k$-mer appears in $s$, we have some information about where it occurs.

\begin{figure} 
	\centering
	\includegraphics[width=0.93\linewidth]{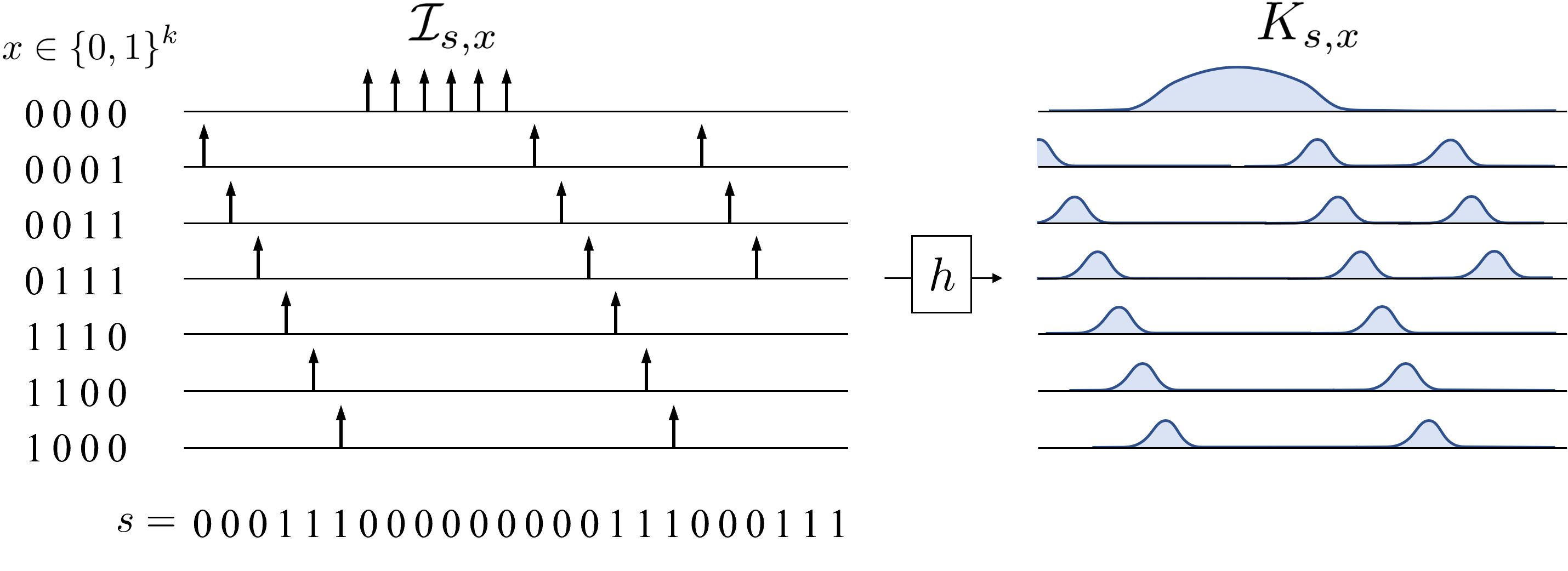}
	\caption{For each $x \in \{0,1\}^k$, $\cI_{s,x}$ indicates the occurrences of $x$ in $s$.
	The density map $K_{s,x}$ can be obtained via $K_{s,x}[i] = \sum_{j=1}^{n-k+1} h(i, j) \mathcal{I}_{s, x}[j]$.
		\label{fig:density}
	}
\end{figure}

For a $k$-mer $x \in \{0, 1\}^k$, let $\mathcal{I}_{s, x} \in \{0, 1\}^{n-k+1}$  be the indicator vector of the occurrences of $x$ in $s$; i.e., 
$\mathcal{I}_{s, x}[j] = \mathbbm{1}\{s_{j:j+k-1} = x \}$,
as illustrated in Figure~\ref{fig:density}.
Notice that recovering the $k$-subword deck can be seen as recovering $\sum_j \cI_{s,x}[j]$ for each $x\in \{0,1\}^k$.
Also notice that recovering $s$ is equivalent to recovering $\mathcal{I}_{s,x}$ for all $x \in \{0,1\}^k$.
A $k$-mer density map can be obtained by computing 
\al{
K_{s,x}[i] & = \sum_{j=1}^{n-k+1} h(i, j) \mathcal{I}_{s, x}[j] 
}
for some ``smoothing kernel'' $h(i,j)$, as illustrated in Figure~\ref{fig:density}.
Intuitively, for a given $x$, $K_{s,x}$ gives a coarse indication of the occurrences of $x$ in $s$.
Moreover, if $h$ is such that $\sum_i h(i,j) = 1$ for each $j$, it holds that
\aln{
\sum_{i} K_{s,x}[i] = \sum_{i} \sum_{j} h(i, j) \cI_{s, x}[j] = \sum_{j} \cI_{s,x}[j] \sum_{i} h(i, j) = \sum_{j} \cI_{s,x}[j],
}
which means that the $k$-subword deck $\cS_k(s)$ is a function of $K_{s,x}$, and the density map $K_{s,x}$ can be thought of as a generalization of the $k$-subword deck that provides information about $k$-mer location.

We will focus on a specific choice of $h(i,j)$ that will render $K_{s,x}$ easier to estimate from the traces.
We will let $h(i,j)$ be the probability that a binomial random variable with $j-1$ trials and probability parameter $1-p$ is equal to $i-1$; i.e., $h(i,j) = \binom{j-1}{i-1} (1-p)^{i-1} p^{j-i}$.
This is also the probability that the $j$th bit of $s$ (if not deleted) ends up as the $i$th bit of a trace.
Hence we have
\al{
K_{s,x}[i] = \sum_{j=1}^{n-k+1} \binom{j-1}{i-1} (1-p)^{i-1} p^{j-i} \cI_{s,x}[j].
}
for $i \in \{1, \dots, n-k+1\}.$
Notice that the maximum value of $h(i,j)$ for a fixed $j$ occurs when $i \approx j(1-p)$ so the kernel $h(\cdot,j)$ has its peak shifted to the left and $K_{s,x}$ is a density map of occurrences of $x$ in $s$ shifted to the left.
Operationally, $(1-p)^k K_{s,x}[i]$ is the probability that 
a fully preserved copy of $x$ in $s$ appears in position $i$ on a given trace of $s$.  

We define the $k$-mer density map of $s$ as  $K_s = [K_{s,x} : x \in \{0,1\}^k ]$ (the concatenation of all vectors $K_{s,x}$).
If the $k$-mer density map $K_s$ is known \emph{exactly}, $s$ can also be recovered exactly.
This can be seen by noticing the invertibility of the upper-triangular matrix $F$ that transforms the binary vector $\mathcal{I}_{s,x}$ into the vector $K_{s,x}$ (for a fixed $x$). 
The matrix $F$ is upper triangular with non-zero entries on the main diagonal, which makes it invertible.
While invertible, $F$ is ill-conditioned since some of the entries on the diagonal are close to $0$, making the transformation from $K_{s,x}$ to $\mathcal{I}_{s,x}$ sensitive to noise in  $K_{s,x}$.

We present an algorithm that, given $T$ traces, constructs an estimate $\hat K_s$ for the $k$-mer density map.
Our main result establishes that we can achieve estimation error
\aln{
\| \hat K_s - K_s \|_\infty = \max_{x,i} | \hat K_{s,x}[i] -  K_{s,x}[i] | < \ep
}
using $T = \ep^{-2}\cdot \poly(n)$ traces.
Hence, the density map $K_s$ can be estimated with  maximum error $\ep_n = 1/g(n)$ for $g(n) \in \poly(n)$ using polynomially many traces.
In particular, given a set of candidate source strings $\cA \subset \{0,1\}^n$ such that, for any $s,s' \in \cA$,
\aln{
\| K_{s} - K_{s'} \|_\infty \geq 2\ep,
}
the true source sequence $s \in \cA$ can be recovered with $\ep^{-2} \cdot \poly(n)$ traces.
This adds to the existing literature on classes of strings recoverable/distinguishable with polynomially many traces~\cite{Mcgregor2014,Sudan,Bruck}. 

Since $\cI_{s,x}$ and $K_{s,x}$ are related through an invertible (albeit ill-conditioned) linear transformation $K_{s,x} = F \cI_{s,x}$, the approximate recovery of the $k$-mer density map $\hat K_{s,x}$ suggests  natural reconstruction algorithms for $\cI_{s,x}$, e.g., based on a regularized least squares problem 
\aln{
\min_{\hat \cI_{s,x}} \| \hat K_{s,x} - F \hat{\cI}_{s,x} \|_2^2 + \delta \| \hat \cI_{s,x} \|_2^2,
}
which is a convex program if $\hat \cI_{s,x}$ is allowed to be real-valued.
The solution $\hat \cI_{s,x}$ can then be converted into a reconstructed string $\hat s \in \{0,1\}^n$ through a majority voting across candidate $k$-mers for each position.
Hence, in contrast to much of the theoretical literature on the trace reconstruction problem, the $k$-mer density map leads to new reconstruction approaches.

Our main result relies on a nontrivial estimator for $K_{s,x}$ that simultaneously uses count information for all binary strings $y$ that are supersequences of $x$.
The estimator is obtained by first deriving a recursive formula for $K_{s,x}$, then applying a known result in the combinatorics of strings on the expansion of the recursive formula to obtain a non-recursive formula. An application of McDiarmid's inequality is then used to prove the estimator is successful with high probability. 
To the best of our knowledge, these techniques have not appeared in the trace reconstruction literature, where most recent results have been based on complex analysis~\cite{PeresNazarov, Peres2,chase2, chase, Servedio}.  
Our techniques also lead to an improvement on a previously known upper bound~\cite{Servedio2} on the number of traces needed for reconstructing the $k$-subword deck of $s$ for $p < 0.5$.



\subsection{Related work}




The current best upper bounds for worst-case trace reconstruction \cite{PeresNazarov,Servedio,chase} are obtained using algorithms that consider each pair of possible source strings $y, z  \in \{0, 1\}^n$, 
and decide whether the set of traces looks more like a set of traces from $y$, or more like a set of traces from $z$ (we formalize this shortly).  
Then if there are enough traces, with high probability the true source string $s$ will be the unique string such that the set of traces looks more like it came from $s$ than 
from any other string in $\{0, 1\}^n$.  
Therefore, the trace reconstruction problem is closely related to the trace-distinguishing problem \cite{Sudan,Bruck},
where 
we want to decide from the set of traces whether the source string is either $y$ or $z$ with high probability.  
The best existing upper bound of $\exp(O(n^{1/5}))$ for worst-case trace reconstruction \cite{chase} is proved using the fact that any two strings can be distinguished using $\exp(O(n^{1/5}))$ traces.
It has also been shown that string pairs at constant Hamming distance can be distinguished using $\text{poly}(n)$ traces by McGregor, Price and Vorotnikova \cite{Mcgregor2014}, and separately by Grigorescu, Sudan, and Zhu using different techniques \cite{Sudan}.  It was recently shown that strings at constant Levenshtein distance can be distinguished using $\text{poly}(n)$ traces by Sima and Bruck \cite{Bruck}.

To distinguish between two strings $y$ and $z$ from a set of traces, current state of the art algorithms identify a function $f_{y,z}$ such that $| E[f_{y,z}(\tilde{Y})] - E[f_{y,z}(\tilde{Z})] |$ is sufficiently large where $\tilde{Y}$ denotes a trace of $y$.  
Given $T$ traces $\tilde{S}_1,\dots, \tilde{S}_T$ of a source string $s$, $E[f_{y,z}(\tilde{S})]$ can be estimated as $\frac{1}{T} \sum_{i=1}^T f_{y,z}(\tilde{S}_i)$ and we say that $y$ {\it beats} $z$ if $\frac{1}{T} \sum_{i=1}^T f_{y,z}(\tilde{S}_i)$ is closer to $E[f_{y,z}(\tilde{Y})]$ than to $E[f_{y,z}(\tilde{Z})]$. 
Observe that if $f_{y, z}$ is such that $| E[f_{y,z}(\tilde{Y})] - E[f_{y,z}(\tilde{Z})] |$ is large enough, then assuming the source string is $y$ or $z$, we can distinguish between the two cases given a reasonable number of traces.
If there is a unique string $u$ such that $u$ beats all other strings, then $u$ is output as the reconstruction.


The first such function $f_{y,z}$ introduced by Nazarov and Peres \cite{PeresNazarov},  and independently by De, O'Donnell, and Servedio \cite{Servedio}, is simply the value of a single bit in the trace, i.e., $f_{y,z}(\tilde{S}) = \tilde{S}[i]$ for source string $s$ and some index $i$ that depends on $y$ and $z$ (\cite{holenstein2008trace} also uses a single bit approach).
An argument based on an existing result on complex-valued polynomials was used to show that for any pair of strings $y, z$, there is some index $i_{y, z}$ such that $|E[\tilde{Y}[i_{y, z}]] - E[\tilde{Z}[i_{y, z}]]| \geq \exp{(-O(n^{1/3}))}$.  Using this result along with a standard concentration inequality, the upper bound of $\exp{(O(n^{1/3}))}$ traces is obtained.
This choice of $f_{y,z}$ is known in the literature as a single bit statistic.  The current best upper bound by Chase \cite{chase} picks a more complicated function $f_{y,z}$ that involves multiple bits, and proves a novel result on complex polynomials in order to prove a gap of $|E[f_{y,z}(\tilde{Y})] - E[f_{y,z}(\tilde{Z})]| \geq \exp{(-\tilde{O}(n^{1/5}))}$, which yields an upper bound of $\exp{(\tilde{O}(n^{1/5}))}$ traces.
A choice of $f_{y,z}$ that uses multiple bits in combination (e.g. that of Chase \cite{chase}) is known in the literature as a multi-bit statistic.  There are similarities between the approach used in \cite{chase} and the approach used in this paper, and it may be possible to obtain results similar to those in this paper based on the complex analysis techniques explored in \cite{chase, PeresNazarov, Servedio}.

Obtaining lower bounds for the number of traces needed in the trace reconstruction problem amounts to proving a lower bound on the number of traces required to distinguish two strings (trace-distinguishing problem) since any algorithm to solve the trace reconstruction problem can be used to solve the trace-distinguishing problem.
For a string $a$, let $a^i$ denote the string where $a$ is repeated $i$ times.  
The current best lower bound of $\tilde{\Omega}(n^{3/2})$ for trace reconstruction discovered by Chase \cite{chase2} is obtained by analyzing the string pair $(01)^{m} 1 (01)^{m+1} $ and $(01)^{m+1} 1 (01)^{m}$ where $n = 2m+3$.


The trace reconstruction problem has also been considered in the smoothed complexity model by Chen, De, Lee, Servedio, and Sinha~\cite{Servedio2}, in which a worst-case string is passed through a binary symmetric channel before trace generation, and the noise-corrupted string needs to be reconstructed with high probability (recall that a binary symmetric channel flips each bit in the string independently with some fixed probability).  
Chen et al. proved that in the smoothed complexity model, trace reconstruction requires $\text{poly}(n)$ traces.  
This result relies on the simple fact that if there are no repeated substrings of length $k-1$ or greater in the source string $s$, then the $k$-subword deck uniquely determines $s$.
The authors prove that for an arbitrary string, the $(\log n)$-subword deck can be reconstructed with high probability using $\text{poly}(n)$ traces, and prove there will not 
be any repeats in the source string of length at least $(\log n-1)$ 
with high probability after it is passed through a binary symmetric channel, thereby proving that $\text{poly}(n)$ traces suffice for trace reconstruction in the smoothed complexity model.

The bulk of the work done in \cite{Servedio2} is proving that the $(\log n)$-subword deck can be reconstructed with high probability using $\text{poly}(n)$ traces for any $p < 1$.  
In order to prove this, a formula for the number of times a substring $x$ is present in the source string $s$ is derived in terms of the expected number of times a trace contains $x$, and the expected number of times a trace contains supersequences of $x$.
The expected values of these statistics are then estimated from the set of traces in order to estimate the number of times $x$ appears in $s$.  Concentration inequalities are used to prove that the number of times $x$ appears is estimated correctly with high probability. 
Narayanan and Ren \cite{Narayanan} provide a similar result on reconstructing the $(100\log n)$-subword deck for circular strings in $\text{poly}(n)$ time.  
Our result showing that the $(\log n)$-subword deck can be computed using $\text{poly}(n)$ traces  with high probability for $p < 0.5$ was originally proved independently, without knowledge of \cite{Servedio2} and \cite{Narayanan}.


\subsection{Notation}
	
	Strings in this paper are binary and indexed starting from 1. If the index $i$ is negative, $x[i]$ is the $(-i)$th element starting from the right end of $x$. For example, if $s = 1001$, then $s[1]=1$, $s[2] = 0$, $s[-1] = 1$, and $s[-2] = 0$. 
	Let $s \in \{0, 1\}^n$ be the length-$n$ string we are trying to recover. The string $s$ will be called the {\it source string}.
	A {\it trace} of $s$ is denoted by $\tilde{S}$, and is generated by deleting each bit of $s$ independently with probability $p$.
	
	  

	For a given string $x$, we let $|x|$ denote the length of $x$. For a string $a$ and integer $r$, $a^r$ denotes the string formed by concatenating $r$ copies of $a$.
	A {\it subsequence} of $x$ is a string that can be formed by deleting elements from $x$, and a {\it supersequence} of $x$ is a string that can be formed by inserting elements into $x$. This is in contrast to a {\it substring} of $x$, which is a string that appears in $x$. We let $x[i, j] = (x[i], x[i+1], \dots, x[j])$ be the substring of $x$ the begins at position $i$ and ends at position $j$. 
	For example, if $s = 10101$, then $s[1:3] = 101$ and $s[2:-2] = 010$.
	For two stings $x$ and $y$, the number of ways we can make $|y|-|x|$ deletions on $y$ to form $x$ is denoted by $\binom{y}{x}$ \cite{Lothaire, Diggavi}.
	This is a generalization of the binomial coefficient for strings. Observe that $\binom{y}{x} \leq \binom{|y|}{|x|}$.  
	To simplify our notation, for strings $x, y$, we will also let $\binom{y}{x}' = \binom{y[2:-2]}{x[2:-2]}$.

	\section{Main Results}

\begin{figure}
    \centering
    
\subfloat[\centering  ]{{

    \tikzstyle{every pin}=[fill=white,
	draw=black,
	font=\footnotesize]
	
    \begin{tikzpicture}[scale=0.8]
    \begin{axis} 
    [width=0.46\textwidth,
    height=0.39\textwidth,
    restrict x to domain*=0:.5,
    restrict y to domain*=0:500,
    title ={$\beta_c(p)$ versus $p$},
    legend cell align={left}, 
    ylabel = $\beta_c(p)$, 
    xlabel = $p$,
    xmin = 0,
    xmax = 0.5,
    ymin = 0,
    yticklabel style={
      /pgf/number format/precision=3,
      /pgf/number format/fixed},
    ]
    
    \addplot [
    line width=1pt,
    color=red!80!black
    ] table[x index=0,y index=1] {data/plt1_0p5.dat};
    \addlegendentry{$c = 0.5$}
    
    \addplot [
    line width=1pt,
    color=green!80!black
    ] table[x index=0,y index=1] {data/plt1_1.dat};
    \addlegendentry{$c = 1$}

    \addplot [
    line width=1pt,
    color=blue!80!black
    ] table[x index=0,y index=1] {data/plt1_2.dat};
    \addlegendentry{$c = 2$}
    \end{axis}   

    \end{tikzpicture}

    }}
    \qquad
    \subfloat[\centering  ]{{

    \tikzstyle{every pin}=[fill=white,
	draw=black,
	font=\footnotesize]
	
    \begin{tikzpicture}[scale=0.8]
    \begin{axis} 
    [width=0.46\textwidth,
    height=0.39\textwidth,
    restrict x to domain*=0:.5,
    restrict y to domain*=0:500,
    title ={Comparison of $k$-subword deck bounds for $c = 1$},
    legend cell align={left}, 
    ylabel = degree of $n$,
    xlabel = $p$,
    xmin = 0,
    xmax = 0.5,
    ymin = 0,
    ymax = 420,
    yticklabel style={
      /pgf/number format/precision=3,
      /pgf/number format/fixed},
    ]
    
    \addplot [
    line width=1pt,
    color=blue!80!black
    ] table[x index=0,y index=1] {data/plt2_new.dat};
    \addlegendentry{new}
    
    \addplot [
    line width=1pt,
    color=red!80!black
    ] table[x index=0,y index=1] {data/plt2_existing.dat};
    \addlegendentry{existing}

    \end{axis}   

    \end{tikzpicture}    
    
    }}

    \caption{(a) Plot of $\beta_c(p)$ for various $c$.  Observe that as $p$ increases, the algorithm requires more traces to achieve the same level of performance.  Similarly, as $c$ increases, more traces are needed.  For all values of $c$, the limit of $\beta_c(p)$ as $p \to 0$ is equal to $3$. 
    (b) Plot of exponent in (\ref{eq:thm2}) versus plot of exponent in 
    (\ref{eq:Servedio_bound}) for $c = 1$.  Observe that our bound is significantly tighter than the existing bound.
    }
    \label{fig:degree}
\end{figure}
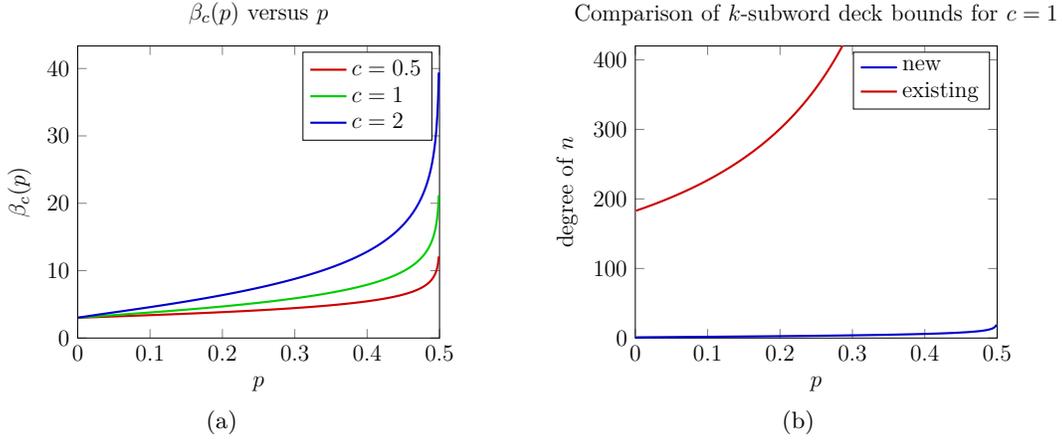

Let $s \in \{0, 1\}^n $ denote the source string and $x \in \{0, 1\}^{k}$ denote the target $k$-mer, whose density $K_{s,x}$  we wish to estimate.
To simplify the notation, we fix a constant $c > 0$ and define 
\al{
f_c(n) = \frac{\left(1 + 2 n^{\alpha_c(p)}  \right)^2}{2 n^{2 c\log(1-p)-1}}
\quad \text{ and } \quad
\alpha_c(p) = 
    1 + c \log\left(\frac{1-p}{p}\right) + \frac{c H(1- \frac{p}{1-p})+c\log\left(\frac{p}{1-p}\right)}{1- \frac{p}{1-p}}
}
where $H(\cdot)$ is the binary entropy function.
The function $f_c(n)$ can be upper bounded by a polynomial of degree $\beta_c(p) = 2\alpha_c(p) - 2c\log(1-p) + 1$, which can be numerically computed as shown in Figure~\ref{fig:degree}(a).
The following theorem is proved in Section~3. 


\begin{theorem} \label{thm:multibit}
    Suppose $p < 0.5$ and $k = c\log n$.
    Given
    $\log\left(\frac{2}{\delta}\right) \cdot \epsilon_n^{-2} \cdot f_c(n)
    $
    traces, an estimator $\hat K_{s,x}[i]$ for the $i$th entry of $K_{s,x}$ can be constructed so that $|\hat K_{s,x}[i] - K_{s,x}[i]| < \ep_n$ with probability $1-\delta$.
    Moreover, given $\log\left(\frac{2n^{1 + c\log 2}}{\delta}\right) \cdot \epsilon_n^{-2} \cdot f_c(n)$ traces, 
    an estimator for the entire density map $\hat K_s$ can be constructed so that $\|\hat K_s - K_s \|_\infty < \ep_n$ with probability $1-\delta$. 
\end{theorem}


In particular, Theorem~\ref{thm:multibit} implies that for $p < 0.5$ and $\epsilon_n = 1/g(n)$ where $g(n) \in  \text{poly}(n)$, all entries of the $(c\log n)$-mer density map $K_{s,x}$ can be estimated with error at most $\epsilon_n$ using $\poly(n)$ traces (and $\poly(n)$ time as discussed in Section 3).
Theorem~\ref{thm:multibit} also implies that the trace reconstruction problem restricted to a set of binary strings with a bounded minimum density map distance can be solved with $\poly(n)$ traces.

\begin{cor}
    \label{cor:distinguish_general}
    Suppose $p < 0.5$ and $k = c\log n$,
    and let $\cA \subset \{0,1\}^n$ be such that, for any $s,s' \in \cA$, 
    \aln{
    \| K_s - K_{s'} \|_\infty \geq 2\ep_n. 
    }
    Given $\log\left(\frac{2n^{1 + c\log 2}}{\delta}\right) \cdot \epsilon_n^{-2} \cdot f_c(n)$ traces from some source string $s \in \cA$, $s$ can be correctly identified with probability $1-\delta$.
\end{cor}




Consequently, the trace reconstruction problem restricted to a set of binary strings with minimum density map distance $1/g(n)$ where $g(n) \in \poly(n)$ can be solved with $\poly(n)$ traces.
We have not been able to find a pair of strings $s$ and $s'$ so that $\| K_{s} - K_{s'} \|_\infty = o(1/g(n))$ for any $g(n) \in \poly(n)$, and to the best of our knowledge, no such example of $s$ and $s'$ is known. 


Recall that, from \cite{Servedio2}, one can recover the $(c\log n)$-subword deck using $\text{poly}(n)$ traces with high probability.
A pair of strings $s,s'$ can be distinguished based on their $(c\log n)$-subword deck alone if and only if their $(c\log n)$-subword decks are distinct, which is equivalent to requiring 
\aln{
|N_{s,x} - N_{s',x}| = |\Vert K_{s, x}\Vert_1 - \Vert K_{s', x}\Vert_1| \geq 1
}
for some $x \in \{0,1\}^{c \log n}$, since $N_{s,x} = \| K_{s,x} \|_1$.
In contrast, our main result implies that 
as long as $ |\Vert K_{s, x}\Vert_1 - \Vert K_{s', x}\Vert_1| \geq 1/\text{poly}(n)$ for some $x$, $s$ and $s'$ can be distinguished with $\poly(n)$ traces, since 
due to the equivalence of $\ell_\infty$ and $\ell_1$ norms and the reverse triangle inequality,
\begin{align}
    \Vert K_{s, x} - K_{s', x} \Vert_\infty \geq \frac{1}{n} \Vert K_{s, x} - K_{s', x} \Vert_1 \geq \frac{1}{n} |\Vert K_{s, x}\Vert_1 - \Vert K_{s', x}\Vert_1|.
\end{align}
This further establishes the $k$-mer density map as a generalization of the $k$-subword deck.  

A special case of Corollary \ref{cor:distinguish_general} with an explicit condition is given below and proved in the appendix. 
\begin{cor}
    \label{cor:distinguish}
    Suppose $p < 0.5$ and $k = c\log n$.
    If the strings $s, s'$ are such that $x \in \{0, 1\}^{k}$ begins at position $i$ in $s$ and $x$ does not appear in $s'$ at an index in the range $[i - f(n), i + f(n)]$ for $f(n) = \Omega(n^a)$ where $a > 0.5$, then $s$ can be distinguished from $s'$ with high probability using $\poly(n)$ traces.
\end{cor}

Corollaries \ref{cor:distinguish_general} and \ref{cor:distinguish} are an addition to the  literature on conditions for distinguishing strings from traces, which includes the facts that strings at constant Hamming distance can be distinguished using $\text{poly}(n)$ traces \cite{Mcgregor2014, Sudan}, and strings at constant Levenshtein distance can be distinguished using $\text{poly}(n)$ traces \cite{Bruck}.  
Observe that there are string pairs that our results immediately prove are poly-trace distinguishable that the results of  \cite{Mcgregor2014, Sudan, Bruck, Servedio2} do not. 
Consider the string pair 
$s = 0^{m} 
  1^{\log(m)}  0^{1.1m}$  and $s' = 0^{1.1m}  1^{\log(m)}  0^{m}$ where $n = 2.1m + \log(m)$. 
Observe that $s, s'$ do not have constant Hamming or Levenshtein distance, and have the same $c \log(n)$ spectrum for any constant $c$, so previous results \cite{Mcgregor2014, Sudan, Bruck, Servedio2} do not apply.




\vspace{2mm}
\noindent {\bf Improved upper bound for $k$-subword deck reconstruction: \;}
Using our proof technique for estimating $K_{s,x}$, we also give a novel proof\footnote{This proof was discovered independently of~\cite{Servedio2} in 2021.} that the $(c\log n)$-subword deck can be reconstructed using $\text{poly}(n)$ traces for $p < 0.5$, which yields an improved upper bound on the required number of traces compared to the analysis of the algorithm for $p < 0.5$ in \cite{Servedio2}.  


\begin{theorem} \label{thm:kdeck_improvement}
    For $p < 0.5$, we can reconstruct the $(c\log n)$-subword deck of any source string $s \in \{0,1\}^n$ from
    \al{
     \tilde{O}\left( n^{1 + c \left(\frac{ (1-p)H(1- p/(1-p)) + p\log(p/(1-p))}{1/2 - p} + 2\log\left(\frac{1}{1-p}\right) \right)}  \right) \label{eq:thm2}
    }
    traces in $\poly(n)$ time with high probability. 
\end{theorem}
In contrast, the analysis in \cite{Servedio2} proves that 
\al{\tilde{O}\left(n^{4 + 12c \left(\frac{e^2}{1/2 - p}\right)    + c\log(4)}\right) \label{eq:Servedio_bound}}
traces are sufficient for this task. 
Comparing the exponents for $p < 0.5$, we prove in the appendix that
\al{
    & 1 + c \left(\frac{ (1-p)H(1- p/(1-p)) + p\log(p/(1-p))}{1/2 - p} + 2\log\left(\frac{1}{1-p}\right) \right) \nonumber
    \\& < 4 + 12c \left(\frac{e^2}{1/2 - p}\right)    + c\log(4) \label{eq:kdeck_degree_relation}
}
which shows that asymptotically, our upper bound is tighter for any $c$ and any $p < 0.5$.  See Figure \ref{fig:degree}(b) for a plot showing the comparison.
In particular, for $c$ close to zero, \eqref{eq:thm2} is close to linear in $n$, nearly matching the following lower bound.
	\begin{theorem}
	    For deletion probability $p$ and any source string $s \in \{0, 1\}^n$, we have that $\Omega(n p (1-p))$ traces are necessary for recovering the $g(n)$-subword deck for any function $g$ such that $g(n) \geq 2$.
	\end{theorem}
	
	\begin{IEEEproof}
	    Suppose we want to decide whether the $k$-subword deck of the source string is the $k$-subword deck of $s_1 = 1^{n/2} 0^{n/2}$, or whether it is the $k$-subword deck of $s_2 = 1^{(n/2) + 1} 0^{(n/2) -1}$.  Observe that for any $k \geq 2$, the only length $n$ string that has the $k$-subword deck of $s_1$ is $s_1$ itself, and likewise, the only length $n$ string that has the $k$-subword deck of $s_2$ is $s_2$ itself.  Thus, distinguishing between these two $k$-subword decks is equivalent to distinguishing between $s_1$ and $s_2$.  From Section 4.2 of \cite{batu2004reconstructing}, we know that $\Omega(n p (1-p))$ traces are necessary for distinguishing between $s_1$ and $s_2$.
	\end{IEEEproof}	 
	We did not compare our upper bound on trace complexity to the analysis of the algorithm in \cite{Servedio2} that reconstructs the $(\log n)$-subword deck using $\text{poly}(n)$ traces for $p < 1$ because an explicit upper bound for this algorithm has not appeared in the literature to the best of our knowledge.
    As discussed in Section 4, we also prove in the appendix that the initial algorithm for $k$-subword deck reconstruction presented in \cite{Servedio2} only needs $\text{poly}(n)$ traces for reconstructing the ($\log n$)-subword deck of a string for $p < 0.5$, which was previously unknown.

\section{An Estimator for the $k$-mer Density Map}

In this section, we describe our estimator for the $k$-mer density map, prove Theorem~\ref{thm:multibit}, and discuss the runtime of computing the estimator.
We first introduce some additional notation.
For a string $x$, let $Y_i(x)$ be the set of length-$i$ supersequences of $x$ that have the same first and last bit of $x$.
For example, if $x = 101$, then $Y_4(x) = \{1011, 1101, 1001\}$.

For source string $s$ and $k$-mer $x$, let $P_{s, x}[i] = \Pr( \tilde{S}[i:i+|x|-1] = x )$, i.e., the probability that $x$ appears at position $i$ in a trace $\tilde{S}$ of $s$.  Notice that it is straightforward to estimate $P_{s, x}[i]$ from the set of traces as  $\hat{P}_{s, x}[i] = \frac{1}{T} \sum_{t = 1}^T \mathbbm{1}\{\tilde{S}_t[i:i+|x|-1] = x \}$.   
Recall that the entry in the $k$-mer density map $K_s$ corresponding to the substring $x$ at index $i$ of $s$ is defined as
\al{
K_{s,x}[i] = \sum_{j=i}^n \binom{j-1}{i-1} (1-p)^{i-1} p^{j-i} \mathbbm{1}\{s[j:j+\ell-1] = x \}. \nonumber
}
In order to estimate $K_{s,x}[i]$, we first write it in terms of $P_{s, x}[i]$ and $P_{s, y}[i]$ for all $y$ of length greater than $k$.  This will then allow us to estimate $K_{s,x}[i]$ using the estimates of $P_{s, x}[i]$ and $P_{s, y}[i]$, which can be obtained directly from the set of traces.  
\begin{lemma} \label{lem:estimator}
    For source string $s$ and $k$-mer $x$, we have
    \al{
        K_{s, x}[i] = \frac{1}{(1-p)^k}\left( P_{s, x}[i] - \sum_{\ell = k+1}^n \sum_{y \in Y_\ell (x)} (-1)^{|y| - |x| + 1} P_{s, y}[i] \binom{y}{x}' \left(\frac{p}{1-p}\right)^{\ell-k} \right).
    }    
\end{lemma}



\begin{IEEEproof}
We begin by deriving a recursive formula for $K_{s, x}[i]$.  We first notice that
\al{
    P_{s, x}[i] & = \sum_{\ell = k}^n \sum_{y \in Y_\ell(x)} \binom{y}{x}' p^{\ell - k} (1-p)^k \sum_{j=i}^n \binom{j-1}{i-1} (1-p)^{i-1} p^{j-i} \mathbbm{1}\{s[j:j+\ell-1] = y \} \nonumber
    \\ & = \sum_{\ell = k}^n \sum_{y \in Y_\ell(x)} \binom{y}{x}' p^{\ell - k} (1-p)^k K_{s, y}[i]. \label{eq:P_sx}
}
This follows because, in order for $x$ to appear at position $i$ in a trace, a superstring $y \in Y_\ell(x)$ must appear at position $j \geq i$ in $s$, $\binom{y}{x}'$ bits from $y$ must be deleted, and $\binom{j-1}{i-1}$ bits in front of $y$ must be deleted.
Notice that 
$\binom{y}{x}' p^{\ell - k} (1-p)^k$ is the probability that a copy of $y$ in $s$ becomes $x$ in $\tilde{S}$, and $K_{s, y}[i]$ is the probability that the beginning of a copy of $y$ in $s$ is shifted to position $i$ in $\tilde{S}$.

Notice that, for $\ell = k$, the only term in the summation in \eqref{eq:P_sx} is $(1-p)^k K_{s,x}[i]$.
This allows us to rewrite \eqref{eq:P_sx} as
\al{
    K_{s, x}[i] & = \frac{1}{(1-p)^k} 
    \left( P_{s, x}[i] - \sum_{\ell = k+1}^n \sum_{y \in Y_\ell(x)} \binom{y}{x}' p^{\ell - k} (1-p)^k K_{s, y}[i] \right) \nonumber
    \\ & = \frac{1}{(1-p)^k} 
    \left( P_{s, x}[i] - \sum_{\ell = k+1}^n \sum_{y \in Y_\ell(x)} (1-p)^\ell \binom{y}{x}'  K_{s, y}[i] \left(\frac{p}{1-p}\right)^{\ell - k} \right). \label{eq:recursive_formula}
}
By recursively applying \eqref{eq:recursive_formula} into itself, we write $K_{s, x}[i]$ in terms of $P_{s, x}[i]$ terms.
This yields
\begin{align}
	K_{s, x}[i] = \frac{1}{(1-p)^k} \left(P_{s, x}[i] - \sum_{\ell=k+1}^n \sum_{y \in Y_\ell(x)} P_{s, y}[i] a_{s, x, y} \left(\frac{ p}{1-p}\right)^{\ell-k} \right)
	\label{eq:formula_with_a}
\end{align}
where $a_{s, x, y} \in \mathbb{Z}$ is a constant that depends on $s, x, y$.
Observe that $a_{s, x, y}$ obeys the following recursion:  for $y \in Y_\ell(x)$, we have that 
\begin{align} \label{eq:coeff_recursion}
	a_{s,x,y} = \binom{y}{x}'  - \sum_{k+1 \leq j < \ell} \;\; \sum_{z \in Y_j(x)} a_{s,x,z} \binom{y}{z}'. 
\end{align}
This is because as we expand (\ref{eq:recursive_formula}) 
one step at a time to eventually obtain (\ref{eq:formula_with_a}), we observe that every time we obtain a new term involving $P_{s, y}[i]$ in the expansion  with coefficient $c_y \left(\frac{p}{1-p}\right)^{|y| - k}$, in the next step of the expansion we obtain a term involving $P_{s, z}[i]$ with coefficient $-c_y \binom{z}{y}' \left(\frac{p}{1-p}\right)^{|z| - k}$ for every $z \in \cup_{\ell = |y| + 1}^n Y_\ell(y)$.  One step of the expansion is shown in the appendix to illuminate this argument.
We proceed to prove that for any $s$, $x$, $y$, we have that 
\begin{align}
	a_{s,x,y} = (-1)^{|y| -|x|+1} \binom{y}{x}'. \label{eq:coeff}
\end{align}
We will use the following lemma, which appears as  Corollary (6.3.9) in \cite{Lothaire}.
\begin{lemma} \label{lem:comb_of_words}
    For any two strings $f, g$ over an alphabet $\Sigma$,
    \begin{align}
        \sum_{h} (-1)^{|g| + |h|} \binom{f}{h} \binom{h}{g} = \delta_{f, g}
    \end{align}
    where $\delta_{f, g} = 0$ if $f \neq g$, and $\delta_{f, g} = 1$ if $f = g$.
\end{lemma}

If $|y| - |x| = 1$,  
\eqref{eq:coeff_recursion} implies that $a_{s,x,y} = \binom{y}{x}'$, and (\ref{eq:coeff}) clearly holds.  
Suppose (\ref{eq:coeff}) holds for $|y| - |x| <  m$.  
Then if $|y| - |x| =  m$, we have that
\begin{align}
	a_{s,x,y} & = \binom{y}{x}' - 	\sum_{k < j < \ell} \;\; \sum_{z \in Y_j(x), } a_{s,x,z} \binom{y}{z}'
	\\ & =  \binom{y}{x}' - 	\sum_{k < j < \ell} \;\; \sum_{z \in Y_j(x), } (-1)^{|z| -|x| + 1} \binom{z}{x}' \binom{y}{z}'
	\\ & =  \binom{y}{x}' + 	\sum_{k < j < \ell} \;\; \sum_{z \in Y_j(x), } (-1)^{|z| + |x|} \binom{z}{x}' \binom{y}{z}'
	\\ & = \binom{y}{x}' +  \left( \sum_{k \leq j \leq \ell} \;\; \sum_{z \in Y_j(x), } (-1)^{|z| + |x|} \binom{z}{x}' \binom{y}{z}' \right) - \binom{x}{x}' \binom{y}{x}' - (-1)^{|y| + |x|} \binom{y}{x}' \binom{y}{y}' 
	\\ & = \binom{y}{x}'   - \binom{y}{x}' - (-1)^{|y| + |x|} \binom{y}{x}'  \label{eq:use_of_comb_lem}
	\\ & =  (-1)^{|y| - |x| + 1} \binom{y}{x}'.
\end{align}	
where \eqref{eq:use_of_comb_lem} follows from Lemma \ref{lem:comb_of_words}.
By plugging in this formula for $a_{s,x,y}$ into  \eqref{eq:formula_with_a}, we obtain
\al{
    K_{s, x}[i] = \frac{1}{(1-p)^k}\left( P_{s, x}[i] - \sum_{\ell = k+1}^n \sum_{y \in Y_\ell (x)} (-1)^{|y| - |x| + 1} P_{s, y}[i] \binom{y}{x}' \left(\frac{p}{1-p}\right)^{\ell-k} \right). \nonumber
}

\end{IEEEproof}


Lemma~\ref{lem:estimator} allows us to obtain an unbiased estimator for  $K_{s, x}[i]$ given by
\al{
    \hat{K}_{s, x}[i] = \frac{1}{(1-p)^k}\left( \hat{P}_{s, x}[i] - \sum_{\ell = k+1}^n \sum_{y \in Y_\ell (x)} (-1)^{|y| - |x| + 1} \hat{P}_{s, y}[i] \binom{y}{x}' \left(\frac{p}{1-p}\right)^{\ell-k} \right)
    \label{eq:kdm-estimator}
}
where $\hat{P}_{s, x}[i] = \frac{1}{T} \sum_{t = 1}^T \mathbbm{1}\{\tilde{S}_t[i:i+|x|-1] = x \}$ and $\hat{P}_{s, y}[i]$ is defined analogously.

One way to analyze the performance of our estimator $\hat K_{s,x}[i]$ would be to apply a standard concentration inequality such as the Chernoff bound to each of the terms $\hat{P}_{s, y}[i]$ and use that to bound the error of $\hat{K}_{s, x}[i]$.
However, this yields a suboptimal analysis as we do not need to guarantee the accuracy of each $\hat{P}_{s, y}[i]$ term.
Directly analyzing the accuracy of $\hat{K}_{s, x}[i]$ is more subtle, as $\hat{K}_{s, x}[i]$ is not a sum of independent random variables.
To that end, we apply McDiarmid's inequality to analyze the deviation of $\hat{K}_{s, x}[i]$ from $K_{s, x}[i]$ directly.  

\begin{lemma} \label{lem:density_map_mcdiamids}
For source string $s$, a $k$-mer $x$ with $|x| = c\log n$, and a set of $T$ traces, we have 
\begin{align}
	\Pr \left(|\hat{K}_{s,x}[i] - K_{s,x}[i]| \geq \epsilon \right) & \leq  2\exp \left(-\frac{2T\epsilon^2}{n \left(\frac{1}{n^{c\log(1-p)}}\left( 1 + 2 n^{\alpha_c(p)}  \right)\right)^2}\right).
\end{align}
\end{lemma}
\begin{IEEEproof}
In order to apply McDiarmid's inequality, we will view the estimator $\hat K_{s,x}[i]$ as a function of $Tn$ independent random variables corresponding to the 
indicator random variables that indicate whether a particular bit is deleted in a particular trace.  
To apply McDiarmid's inequality, we have to upper bound how much the estimator can change by changing the value of one of these indicator random variables.   
Changing the value of the indicator random variable for a particular bit in the $t$th trace $\tilde{S}_t$ changes the estimator by at most
\begin{align}
    b & \leq  \frac{1}{T} \frac{1}{(1-p)^k}\left( 1 + \sum_{\ell = k+1}^n 2 \max_{y \in Y_\ell (x)} \binom{y}{x}' \left(\frac{p}{1-p}\right)^{\ell-k}\right)
    \\ & \leq \frac{1}{T} \frac{1}{(1-p)^k}\left(1 + \sum_{\ell = k+1}^n 2  \binom{\ell-2}{k-2} \left(\frac{p}{1-p}\right)^{\ell-k}\right)
    \\ & \leq \frac{1}{T} \frac{1}{(1-p)^k}\left( 1 + 2 n \max_{\ell \in [k+1, n]}   \binom{\ell-2}{k-2} \left(\frac{p}{1-p}\right)^{\ell-k}\right).
\end{align}
where the first inequality follows because changing a single bit in a single trace can change $\hat{P}_{s, x}[i]$ by at most $1/T$, and can change $\hat{P}_{s, y}[i]$ by at most $1/T$ for exactly two values of $y \in Y_\ell(x)$ each $\ell$.
To analyze $b$ when $k = c \log(n)$ for $c$ constant, and $p < 0.5$ we have that
\begin{align} b & \leq \frac{1}{T} \frac{1}{(1-p)^{c\log(n)}}\left( 1 + 2 n \max_{\ell \in [c\log(n)+1, n]}  \binom{\ell-2}{c\log(n)-2} \left(\frac{p}{1-p}\right)^{\ell-c\log(n)}\right)
\\ & \leq \frac{1}{T} \frac{1}{(1-p)^{c\log(n)}}\left( 1 + 2 n \max_{\ell \in [c\log(n)+1, n]}  \binom{\ell}{c\log(n)} \left(\frac{p}{1-p}\right)^{\ell-c\log(n)}\right)
\\ & = \frac{1}{T} \frac{1}{n^{c\log(1-p)}}\left( 1 + 2 n^{1 + c \log(\frac{1-p}{p})}  \max_{\ell \in [c\log(n)+1, n]}  \binom{\ell}{c\log(n)} \left(\frac{p}{1-p}\right)^{\ell}\right).
\end{align}
We have that
\begin{align}
    & \max_{\ell \in [c\log(n)+1, n]}  \binom{\ell}{c\log(n)} \left(\frac{p}{1-p}\right)^{\ell}
    \\ & \leq \max_{\ell \in [c\log(n)+1, n]} 2^{\ell H(c\log(n)/\ell)} \left(\frac{p}{1-p}\right)^{\ell}
    \\ & \leq 2^{ \frac{\log(n^c)}{1- p/(1-p)} H(1- p/(1-p))} \left(\frac{p}{1-p}\right)^{\frac{c \log(n)} {1 - p/(1-p)}}
    \\ & = n^{\frac{c H(1- p/(1-p))}{1- p/(1-p)} + \frac{c\log(p/(1-p))}{1 - p/(1-p)}} 
\end{align}
because the maximizing $\ell$ is given by $\ell^* = \frac{\log(n^c)}{1- p/(1-p)}$ .
This is because $2^{x H(c\log(n)/x)} q^{x}$ is differentiable for $x \in (c\log(n)+1, n)$ where $q = \frac{p}{1-p}$, and $x = \frac{\log(n^c)}{1- q}$ is the only zero of
\begin{align}
    & \frac{d}{dx} 2^{x H(c\log(n)/x)} q^{x} \\ & = 2^{x H(c\log(n)/x)} q^{x} (\log(q) - \log(1 - \log(n^c)/x)),
\end{align}
and is a local maximum.  It is easy to see that $x = \frac{\log(n^c)}{1- q}$ is a local maximum by observing that  derivative is positive at $x = \frac{\log(n^c)}{1- q} - \epsilon$ and negative at
$x = \frac{\log(n^c)}{1- q} + \epsilon$  for $\epsilon > 0$ small enough.
We therefore have that we  have that 
\begin{align} 
    b &
    \leq  \frac{1}{T} \frac{1}{n^{c\log(1-p)}}\left( 1 + 2 n^{1 + c \log((1-p)/p)}  n^{\frac{c H(1- p/(1-p))}{1- p/(1-p)} + \frac{c\log(p/(1-p))}{1 - p/(1-p)}} \right)
    \\ & = \frac{1}{T} \frac{1}{n^{c\log(1-p)}}\left( 1 + 2 n^{1 + c \log((1-p)/p) + \frac{c H(1- p/(1-p))}{1- p/(1-p)} + \frac{c\log(p/(1-p))}{1 - p/(1-p)}} \right)
\end{align}    
when $k = c\log(n)$ for $c$ constant and $p < 0.5$.
Let
\begin{align}
    \alpha_c(p) = 1 + c \log((1-p)/p) + \frac{c H(1- p/(1-p))}{1- p/(1-p)} + \frac{c\log(p/(1-p))}{1 - p/(1-p)}.
\end{align}
Plugging this into McDiarmid's inequality,  we have
\begin{align}
	\Pr \left(|\hat{K}_{s,x}[i] - K_{s,x}[i]| \geq \epsilon \right) & \leq 2\exp \left(-\frac{2\epsilon^2}{nT  \left(\frac{1}{T} \frac{1}{n^{c\log(1-p)}}\left( 1 + 2 n^{\alpha_c(p)}  \right) \right)^2}\right)
	\\ & = 2\exp \left(-\frac{2T\epsilon^2}{n \left(\frac{1}{n^{c\log(1-p)}}\left( 1 + 2 n^{\alpha_c(p)}  \right)\right)^2}\right).
\end{align}
\end{IEEEproof}
Setting $\delta = \Pr \left(|\hat{K}_{s,x}[i] - K_{s,x}[i]| \geq \epsilon \right)$, we conclude that
\begin{align}
    T = \log(2/\delta) \frac{n}{2\epsilon^2} \left(\frac{1}{n^{c\log(1-p)}}\left(1 + 2 n^{\alpha_c(p)}  \right)\right)^2
\end{align}
traces suffices for recovering $K_{s,x}[i]$ with error less than $ \epsilon$ with probability at least $1 - \delta$.


Observe that we can compute the estimate for $K_{s,x}[i]$  in (\ref{eq:kdm-estimator}) by iterating through all $T$ traces and constructing a linked list for each $\ell \in \{k+1, ..., n\}$ that stores all the length-$\ell$ strings that start at position $i$ in a trace, along with the number of times it is seen in the set of traces.  This requires $O(T^2 n^3 )$ time and $\tilde{O}(T n^3 )$ space because there are at most $n$ substrings $y$ starting at position $i$ in each trace, and $\binom{y}{x}'$ can be computed in $O(n^2)$ time using dynamic programming (see Proposition 6.3.2 in \cite{Lothaire}).  
It is worth noting that the expected runtime would be faster in practice if a hash table is used to store each $y$ seen for each $\ell$.  

In conjunction with our trace complexity analysis, the approach described above for computing the estimator implies that for $x$ of length $c \log(n)$, we can estimate $K_{s,x}[i]$ with error at most $1/\text{poly}(n)$ with high probability in $\text{poly}(n, 1/\epsilon)$ time, and can therefore estimate the $c \log(n)$-density map with maximum error $1/\text{poly}(n)$ with high probability in $\text{poly}(n, 1/\epsilon)$ time.



	\section{A New Bound for the $k$-subword deck Reconstruction Problem}
	
	In this section, we derive an estimator for the $k$-subword deck using the same techniques used to derive the estimator for the $k$-mer density map, and prove Theorem \ref{thm:kdeck_improvement}.
	Let $N_{s, x}$ denote the number of times the $k$-mer $x$ appears in the source string $s$.  When $s$ is clear from context, we write $N_{s, x}$ as $N_x$.  In this section, let $E_{s, x}$ denote the expected number of times that $k$-mer $x$ appears in a trace $\tilde{S}$ of source string $s$. When $s$ is clear from context, we write $E_{s, x}$ as $E_x$. Notice that we can easily estimate $E_x$ from a set of $T$ traces as  $\hat{E}_x = \frac{1}{T}\sum_{i = 1}^T (\text{\# of times $x$ appears as a substring in trace $\tilde{S}_i$})$.

	Similar to the $k$-mer density map estimator, we derive a formula for $N_x$ in terms of $E_x$ and $E_y$ for all supersequences $y$ of $x$.  This allows us to estimate $N_x$ directly from the set of traces.  Our formula for $N_x$ is given in the following lemma.  Refer to the beginning of the previous section for the definition of $Y_i(x)$.
	
	\begin{lemma} \label{lem:kdeck-estimator}
	    For source string $s \in \{0, 1\}^n$, deletion probability $p \in [0, 1)$, and $x \in \{0, 1\}^k$, we have  
	    \begin{align}
	        N_x = \frac{1}{(1-p)^k}\left( E_x - \sum_{i = k+1}^n \sum_{y \in Y_i (x)} (-1)^{|y| - |x| + 1} E_y  \binom{y}{x}' \left(\frac{p}{1-p}\right)^{i-k} \right).
	    \end{align}
	\end{lemma}	
	
	\begin{IEEEproof}
	Observe that 
	\begin{align}
	E_x  & = \sum_{i=k}^n \sum_{y \in Y_i(x)} N_{y} \binom{y}{x}' p^{i-k} (1-p)^k \nonumber
	\\ & = \sum_{i=k}^n \sum_{y \in Y_i(x)} (1-p)^i N_{y} \binom{y}{x}' \left(\frac{p}{1-p}\right)^{i-k} 
    \end{align}
    by linearity of expectation applied to all ways $x$ can be formed from a substring $y$ in $s$.  This can be rewritten as the following recursive formula for $N_x$  
	\begin{align}
	N_x = \frac{1}{(1-p)^k}\left(E_x - \sum_{i=k+1}^n \sum_{y \in Y_i(x)} (1-p)^i N_{y} \binom{y}{x}' \left(\frac{p}{1-p}\right)^{i-k} \right).
    \end{align}
	 This formula is identical to the recursive formula for the $k$-mer density map in  (\ref{eq:recursive_formula}) with $K_{s, x}[i]$ replaced by $N_x$ and $E_{s, y}[i]$ replaced by $E_y$.  Therefore, the same steps can be taken to obtain a non-recursive formula for $N_x$, which is given by 
    \begin{align}
	        N_x = \frac{1}{(1-p)^k}\left( E_x - \sum_{i = k+1}^n \sum_{y \in Y_i (x)} (-1)^{|y| - |x| + 1} E_y  \binom{y}{x}' \left(\frac{p}{1-p}\right)^{i-k} \right). \nonumber
	 \end{align}
	\end{IEEEproof}
    Our estimator is given by
    \begin{align}
        \hat{N}_x = \frac{1}{(1-p)^k}\left( \hat{E}_x - \sum_{i = k+1}^n \sum_{y \in Y_i (x)} (-1)^{|y| - |x| + 1} \hat{E}_y  \binom{y}{x}' \left(\frac{p}{1-p}\right)^{i-k} \right).
	\end{align}
	which we then round to the nearest integer.  Observe that if the error in the estimate of $N_x$ is less than $0.5$, then we will estimate $N_x$ correctly after rounding. 
    We can implement this estimator efficiently using a similar approach to the one described in Section 3,  allowing us to compute the $k$-subword deck in  $\text{poly}(n, 1/\epsilon)$ time and space for $k = c \log(n)$.

Chen, De, Lee, Servedio, and Sinha \cite{Servedio2} derive a formula for $N_x$ that is equivalent to ours, but written in a different form.  Their formula is given by 
\al{
    N_x = \frac{1}{(1-p)^k} \sum_{\substack{\alpha \in \mathbb{Z}_{\geq 0}^{k-1} \\ |\alpha| \leq n-k}} E \left[\#(x[1] *^{\alpha[1]} x[2] *^{\alpha[2]} x[3] \cdot \cdot \cdot  x[k-1] *^{\alpha[k-1]} x[k], \; \tilde{S})\right] \left(\frac{-p}{1-p}\right)^{|\alpha|} \label{eq:Servedio_polynomial}
}
where $*$ is the wildcard symbol and $\#(x[1] *^{\alpha[1]} x[2] *^{\alpha[2]} x[3] \cdot \cdot \cdot  x[k-1] *^{\alpha[k-1]} x[k], \; \tilde{S})$ is the number of times a substring of the form $x[1] *^{\alpha[1]} x[2] *^{\alpha[2]} x[3] \cdot \cdot \cdot  x[k-1] *^{\alpha[k-1]} x[k]$ appears in a trace $\tilde{S}$ of $s$, and $|\alpha| = \sum_{i=1}^{k-1} |\alpha[i]|$ for $\alpha \in \mathbb{Z}_{\geq 0}^{k-1}$.
In order to prove this formula, the authors define a  complex polynomial that is a generalization of $N_x$, perform a Taylor expansion on the polynomial, and relate the partial derivatives of the polynomial to expected values of functions of traces of $s$. The formula above is then a special case of this more general result.   
Our significantly shorter proof is very different from theirs in that we first derive a recursive estimator of $N_x$ and then use structural knowledge of $\binom{y}{x}'$ to prove its equivalence to the polynomial above.  

The initial algorithm presented in \cite{Servedio2} is to use (\ref{eq:Servedio_polynomial}) directly for estimation of $N_x$. For $p < 0.5$, the authors bound the error of the estimate of each $ E \left[\#(x[1] *^{\alpha[1]} x[2] *^{\alpha[2]} x[3] \cdot \cdot \cdot  x[k-1] *^{\alpha[k-1]} x[k], \; \tilde{S})\right]$ term using a Chernoff bound and the fact that $\#(x[1] *^{\alpha[1]} x[2] *^{\alpha[2]} x[3] \cdot \cdot \cdot  x[k-1] *^{\alpha[k-1]} x[k], \; \tilde{S}) \in [0, n]$.  This analysis of the  algorithm yields an upper bound of $n^{O(k)} \cdot \log(1/\delta)$ traces.  For $k = c \log(n)$, this is superpolynomial  in $n$.  In the appendix, we employ McDiarmid's inequality to prove that $\text{poly}(n)$ traces suffice for reconstructing $N_x$ from (\ref{eq:Servedio_polynomial}) with high probability.

In the improved algorithm \cite{Servedio2} for $p < 0.5$ that uses $\text{poly}(n)$ traces, the authors estimate $N_x$ using only estimates of $ E \left[\#(x[1] *^{\alpha[1]} x[2] *^{\alpha[2]} x[3] \cdot \cdot \cdot  x[k-1] *^{\alpha[k-1]} x[k], \; \tilde{S})\right]$ for $|\alpha| \leq d+k$ where $d = \frac{e^2}{1/2 - p} \left( k\log \left(\frac{e^2}{1/2 - p}\right) + \log(n) \right)$.  The truncated formula they use is 
\al{
    N_x = \frac{1}{(1-p)^k} \sum_{\substack{\alpha \in \mathbb{Z}_{\geq 0}^{k-1} \\ |\alpha| \leq d}} E \left[\#(x[1] *^{\alpha[1]} x[2] *^{\alpha[2]} x[3] \cdot \cdot \cdot  x[k-1] *^{\alpha[k-1]} x[k], \; \tilde{S})\right] \left(\frac{-p}{1-p}\right)^{|\alpha|}. \label{eq:Servedio_truncated_polynomial}
}
The authors then prove that the sum of the terms in (\ref{eq:Servedio_polynomial}) for strings $y$ of length $>d+k$ is small, and again bound the error of the estimate of each $ E \left[\#(x[1] *^{\alpha[1]} x[2] *^{\alpha[2]} x[3] \cdot \cdot \cdot  x[k-1] *^{\alpha[k-1]} x[k], \; \tilde{S})\right]$ for $|\alpha| \leq d$ using a Chernoff bound and the fact that $\#(x[1] *^{\alpha[1]} x[2] *^{\alpha[2]} x[3] \cdot \cdot \cdot  x[k-1] *^{\alpha[k-1]} x[k], \; \tilde{S}) \in [0, n]$. 
In specific, they prove Lemma \ref{lem:trancation_bound} to upper bound the magnitude of the sum of the truncated terms in (\ref{eq:Servedio_polynomial}), and prove that the truncated polynomial (\ref{eq:Servedio_truncated_polynomial}) can be estimated with error at most $0.2$ with high probability.  This leads to an estimator of (\ref{eq:Servedio_polynomial}) with error at most $0.3$ with high probability, which is sufficient for estimating $N_x$ correctly after rounding.
\begin{lemma} \label{lem:trancation_bound}
    For $p < 1/2$, and $d = \frac{e^2}{1/2 - p} \left( k\log \left(\frac{e^2}{1/2 - p}\right) + \log(n) \right)$, 
    \al{
        \left| \frac{1}{(1-p)^k} \sum_{\substack{\alpha \in \mathbb{Z}_{\geq 0}^{k-1} \\ |\alpha| > d}} E \left[\#(x[1] *^{\alpha[1]} x[2] *^{\alpha[2]} x[3] \cdot \cdot \cdot  x[k-1] *^{\alpha[k-1]} x[k], \; \tilde{S})\right] \left(\frac{-p}{1-p}\right)^{|\alpha|}\right|  \leq 0.1.
    }
\end{lemma}

The number of traces used by the truncated algorithm to reconstruct $N_x$ with probability at least $1-\delta$ for $p < 1/2$ 
is proved to be $O((M^2 2^k)^2) \log\left(\frac{M}{\delta}\right)$, and it is proved that  $M \leq n \left(\frac{e^2}{1/2 - p}\right)^{3k}$.  It follows that number of traces used in the analysis of \cite{Servedio2} is
\begin{align}
    & O\left( \left( \left(n \left(\frac{e^2}{1/2 - p}\right)^{3k} \right)^2 2^k\right)^2\right) \log\left(\frac{n \left(\frac{e^2}{1/2 - p}\right)^{3k}}{\delta}\right) \nonumber
    \\ & = O\left( n^4  \left(\frac{e^2}{1/2 - p}\right)^{12k}  4^k \right) \log\left(\frac{n \left(\frac{e^2}{1/2 - p}\right)^{3k}}{\delta}\right).
\end{align}
For $k = c \log(n)$, the number of traces used is
\begin{align}
    & O\left( n^4  \left(\frac{e^2}{1/2 - p}\right)^{12c\log(n)}  4^{c\log(n)} \right) \log\left(\frac{n \left(\frac{e^2}{1/2 - p}\right)^{3c\log(n)}}{\delta}\right) \nonumber
    \\ & = O\left( n^4  n^{12c \left(\frac{e^2}{1/2 - p}\right)}   n^{c\log(4)} \right) \log\left(\frac{n \cdot n^{3c\log\left(\frac{e^2}{1/2 - p}\right)}}{\delta}\right) \nonumber
    \\ & = O\left( n^{4 + 12c \left(\frac{e^2}{1/2 - p}\right)    + c\log(4)} \right) \log\left(\frac{n^{1 + 3c\log\left(\frac{e^2}{1/2 - p}\right)}}{\delta}\right).
\end{align}

We show that by using McDiarmid's to analyze sample complexity, we can significantly improve the upper bound on the number of samples needed by the truncated estimator for reconstructing the $k$-subword deck. This result immediately yields Theorem \ref{thm:kdeck_improvement}.  
\begin{lemma} \label{lem:truncated_est_analysis}
    For deletion probability $p < 0.5$, and any source string $s \in \{0,1\}^n$, we can reconstruct $N_x$ for any $k$-mer $x$ where $k = c\log n$ with probability $1-\delta$ using
    \al{
        O \left(\log^4(n) \cdot n^{1 + c \left(\frac{ (1-p)H(1- p/(1-p)) + p\log(p/(1-p))}{1/2 - p} + 2\log\left(\frac{1}{1-p}\right) \right)}\right)\log(1/\delta) 
    }
    traces.
\end{lemma}
\begin{IEEEproof}
    Recall that the truncated estimator from \cite{Servedio2} paper written in our form is given by 
    \begin{align}
        \hat{N}_x = \frac{1}{(1-p)^k}\left( \hat{E}_x - \sum_{i = k+1}^d \sum_{y \in Y_i (x)} (-1)^{|y| - |x| + 1} \hat{E}_y  \binom{y}{x}' \left(\frac{p}{1-p}\right)^{i-k} \right) \nonumber
	\end{align}
	where 
	\begin{align}
	    d = \frac{e^2}{1/2 - p} \left( k \ln \frac{e^2}{1/2 - p} + \ln n \right). \nonumber
	\end{align}
The set of independent random variables we use in McDiarmid's inequality is the set of $Tn$ indicator random variables that indicate whether a particular bit is deleted in a particular trace.  To apply McDiarmid's inequality, we have to upper bound how much the estimator can change by changing the value of one of the indicator random variables.   
Changing the value of the indicator random variable for a particular bit in the $t$th trace $\tilde{S}_t$ changes the estimator by at most
	\begin{align}
	    b & \leq  \frac{1}{T} \frac{1}{(1-p)^k}\left( \min(k, n-k+1) + \sum_{i = k+1}^d 2 \min(i, n-i+1) \max_{y \in Y_i (x)} \binom{y}{x}' \left(\frac{p}{1-p}\right)^{i-k}\right) \nonumber
	    \\ & \leq \frac{1}{T} \frac{1}{(1-p)^k}\left( k + \sum_{i = k+1}^d 2 i  \binom{i-2}{k-2} \left(\frac{p}{1-p}\right)^{i-k}\right) \nonumber
	    \\ & \leq \frac{1}{T} \frac{1}{(1-p)^k}\left( k + 2 d \max_{i \in [k+1, d]} i  \binom{i-2}{k-2} \left(\frac{p}{1-p}\right)^{i-k}\right) \nonumber
	    \\ & \leq \frac{1}{T} \frac{1}{(1-p)^k}\left( k + 2 d^2 \max_{i \in [k+1, d]}  \binom{i-2}{k-2} \left(\frac{p}{1-p}\right)^{i-k}\right)
	\end{align}
	To analyze $b$ when $k = c \log(n)$ for $c$ constant, and $p < 0.5$ we have that
    \begin{align} b & \leq \frac{1}{T} \frac{1}{(1-p)^{c\log(n)}}\left( c\log(n) + 2 d^2 \max_{i \in [c\log(n)+1, d]}  \binom{i-2}{c\log(n)-2} \left(\frac{p}{1-p}\right)^{i-c\log(n)}\right) \nonumber
    \\ & \leq \frac{1}{T} \frac{1}{(1-p)^{c\log(n)}}\left( c\log(n) + 2 d^2 \max_{i \in [c\log(n)+1, d]}  \binom{i}{c\log(n)} \left(\frac{p}{1-p}\right)^{i-c\log(n)}\right) \nonumber
    \\ & = \frac{1}{T} \frac{1}{n^{c\log(1-p)}}\left( c\log(n) + 2d^2 n^{c \log(\frac{1-p}{p})}  \max_{i \in [c\log(n)+1, d]}  \binom{i}{c\log(n)} \left(\frac{p}{1-p}\right)^{i}\right).
	\end{align}
	We have that
	\begin{align}
	    & \max_{i \in [c\log(n)+1, n]}  \binom{i}{c\log(n)} \left(\frac{p}{1-p}\right)^{i}
	    \\ & \leq \max_{i \in [c\log(n)+1, n]} 2^{i H(c\log(n)/i)} \left(\frac{p}{1-p}\right)^{i} \label{eq:kdeck_trunc_upper_bound_lemma}
	    \\ & \leq 2^{ \frac{\log(n^c)}{1- p/(1-p)} H(1- p/(1-p))} \left(\frac{p}{1-p}\right)^{\frac{c\log(n)} {1 - p/(1-p)}}
	    \\ & = n^{\frac{c H(1- p/(1-p))}{1- p/(1-p)} + \frac{c\log(p/(1-p))}{1 - p/(1-p)}} 
	\end{align}
	where the maximizer in (\ref{eq:kdeck_trunc_upper_bound_lemma}) is given by $i = \frac{\log(n^c)}{1- p/(1-p)}$ as proved in Lemma \ref{lem:density_map_mcdiamids}.
    We therefore have that 
        \begin{align} b &
        \leq  \frac{1}{T} \frac{1}{n^{c\log(1-p)}}\left( c\log(n) + 2 d^2 n^{c \log((1-p)/p)}  n^{\frac{c H(1- p/(1-p))}{1- p/(1-p)} + \frac{c\log(p/(1-p))}{1 - p/(1-p)}} \right) \nonumber
        \\ & = \frac{1}{T} \frac{1}{n^{c\log(1-p)}}\left( c\log(n) + 2 d^2 n^{c \log((1-p)/p) + \frac{c H(1- p/(1-p))}{1- p/(1-p)} + \frac{c\log(p/(1-p))}{1 - p/(1-p)}} \right)
	\end{align}    
    when $k = c\log(n)$ for $c$ constant and $p < 0.5$.
    Let
    \begin{align}
        \gamma_c(p) = c \log((1-p)/p) + \frac{c H(1- p/(1-p))}{1- p/(1-p)} + \frac{c\log(p/(1-p))}{1 - p/(1-p)}
    \end{align}
	Due to Lemma \ref{lem:trancation_bound}, it suffices to estimate $N_x$ with error at most $1/5$. Using McDiarmid's inequality, and setting $\epsilon = 1/5$,  we have
	\begin{align}
		\Pr \left(\hat{N}_x - N_x \geq \frac{1}{5} \right) & \leq \exp \left(-\frac{2\left( \frac{1}{5}\right)^2}{nT  \left(\frac{1}{T} \frac{1}{n^{c\log(1-p)}}\left( c\log(n) + 2 d^2 n^{\gamma_c(p)}  \right) \right)^2}\right) \nonumber
		\\ & = \exp \left(-\frac{T}{\frac{25}{2}n \left(\frac{1}{n^{c\log(1-p)}}\left( c\log(n) + 2 d^2 n^{\gamma_c(p)}  \right)\right)^2}\right)
    \end{align}
    Setting $\delta = \Pr \left(\hat{N}_x - N_x \geq \frac{1}{5} \right)$, we have
    \begin{align}
        T = \log(1/\delta) \frac{25}{2} n \left(\frac{1}{n^{c\log(1-p)}}\left( c\log(n) + 2 d^2 n^{\gamma_c(p)}  \right)\right)^2
    \end{align}
    traces suffices for recovering $N_x$ with probability at least $1 - \delta$.

The number of traces needed to recover $N_x$ with probability at least $1- \delta$ for $p < 1/2$ using the truncated estimator is
\begin{align}
    & \log(1/\delta) \frac{25}{2} n \left(\frac{1}{n^{c\log(1-p)}}\left( c\log(n) + 2 d^2 n^{\gamma_c(p)}  \right)\right)^2 \nonumber
    \\ & = O \left(d^4 n^{1 + 2 \left(\gamma_c(p) + c\log\left(\frac{1}{1-p}\right) \right)}\right)\log(1/\delta) 
\end{align}
traces where
\begin{align}
    \gamma_c(p) & = c \log((1-p)/p) + \frac{c H(1- p/(1-p))}{1- p/(1-p)} + \frac{c\log(p/(1-p))}{1 - p/(1-p)} \nonumber
    \\ & =  \frac{c( -1 + p/(1-p))\log(p/(1-p))}{1 - p/(1-p)} + \frac{c H(1- p/(1-p))}{1- p/(1-p)} + \frac{c\log(p/(1-p))}{1 - p/(1-p)} \nonumber
    \\ & = \frac{c (1-p)H(1- p/(1-p))}{1- 2p} + \frac{ cp\log(p/(1-p))}{1- 2p} \nonumber
    \\ & = \frac{c (1-p)H(1- p/(1-p)) + cp\log(p/(1-p))}{1- 2p}.
\end{align}
Thus, the number of traces needed by our algorithm is upper bounded by 
\begin{align}
    & O \left(d^4 n^{1 + 2 \left(\frac{c (1-p)H(1- p/(1-p)) + cp\log(p/(1-p))}{1- 2p} + c\log\left(\frac{1}{1-p}\right) \right)}\right)\log(1/\delta) \nonumber
    \\ & = O \left(d^4 n^{1 + 2c \left(\frac{ (1-p)H(1- p/(1-p)) + p\log(p/(1-p))}{1- 2p} + \log\left(\frac{1}{1-p}\right) \right)}\right)\log(1/\delta) \nonumber
    \\ & = O \left(d^4 n^{1 + c \left(\frac{ (1-p)H(1- p/(1-p)) + p\log(p/(1-p))}{1/2 - p} + 2\log\left(\frac{1}{1-p}\right) \right)}\right)\log(1/\delta) 
\end{align}
Plugging in 
\begin{align}
    d = \frac{e^2}{1/2 - p} \left( k \ln \frac{e^2}{1/2 - p} + \ln n \right),
\end{align}
the number of traces is upper bounded by
\al{
    O \left(\log^4(n) \cdot n^{1 + c \left(\frac{ (1-p)H(1- p/(1-p)) + p\log(p/(1-p))}{1/2 - p} + 2\log\left(\frac{1}{1-p}\right) \right)}\right)\log(1/\delta). 
}
\end{IEEEproof}

	\bibliographystyle{ieeetr}
\bibliography{refs}

    \section{Appendix}
    
    \subsection{Proof of Corollary \ref{cor:distinguish}}
    
    Without loss of generality, assume that $x$ occurs at position $\kappa(n) = \Theta(n)$ in $s$ (if $\kappa(n) \neq \Theta(n)$ we can apply the following argument where $s$, $s'$ and $x$ are reversed).   Observe that for $i \in \{1, \dots, n-k+1\},$ 
    \al{
    K_{s,x}[i] & = \sum_{j=1}^{n-k+1} \binom{j-1}{i-1} (1-p)^{i-1} p^{j-i} \cI_{s,x}[j] \nonumber
    \\ & = \sum_{j=1}^{n-k+1} \Pr(\text{Bin}(j-1, 1-p) = i-1) \cI_{s,x}[j] 
    }
    where $\text{Bin}(n, q)$ is a binomial random variable with $n$ trials and success probability $q$.
    In specific, at $i = \lceil(1-p) \kappa(n) \rceil$, we have that 
    \al{
        K_{s, x}[\lceil(1-p) \kappa(n) \rceil ] & \geq \Pr(\text{Bin}(\kappa(n) -1, 1-p) = \lceil(1-p) \kappa(n) \rceil - 1 ) \nonumber
        \\ & = \Pr(\mathcal{N}((1-p) (\kappa(n) - 1), \; p(1-p) (\kappa(n) - 1)) = \lceil(1-p) \kappa(n) \rceil - 1) + o\left(\frac{1}{\sqrt{n}}\right) \nonumber
        \\ & = \frac{1}{\sqrt{2\pi p(1-p)(\kappa(n)-1)}} e^{-\frac{((1-p)(\kappa(n) - 1) - \lceil(1-p) \kappa(n) \rceil + 1  )^2}{p(1-p) (\kappa(n) - 1)}} + o\left(\frac{1}{\sqrt{n}}\right) \nonumber
        \\ & \geq \frac{1}{\sqrt{2\pi p(1-p)(\kappa(n)-1)}} e^{-\frac{(p+1) ^2}{p(1-p) (\kappa(n) - 1)}} + o\left(\frac{1}{\sqrt{n}}\right) \nonumber
        \\ & = \frac{1}{\sqrt{2\pi p(1-p)(\kappa(n)-1)}} e^{-\frac{(p+1) ^2}{p(1-p) (\kappa(n) - 1)}} + o\left(\frac{1}{\sqrt{n}}\right) \nonumber
        \\ & = \Omega \left(\frac{1}{\sqrt{n}} \right)
    }
    by the local limit theorem.
    On the other hand, we have that for some function $g$ such that $g(n) = \Omega(n^a)$ where $a > 0.5$,
    \al{
        K_{s', x}[\lceil(1-p) \kappa(n) \rceil ] & \leq n \max\bigg( \Pr \left(\text{Bin}(\kappa(n) - g(n) -1, 1-p) = \lceil(1-p) \kappa(n) \rceil - 1 \right), \nonumber  \\ & \quad \Pr(\text{Bin}(\kappa(n) + g(n) -1, 1-p) = \lceil(1-p) \kappa(n) \rceil - 1 ) \bigg) \nonumber
        \\ & \leq n \Pr(|\text{Bin}(\kappa(n) - g(n) -1, 1-p) - (1-p)(\kappa(n) - g(n) -1)| > (1-p)g(n) +p) \nonumber
        \\ & \leq n 2\exp\left(-\frac{((1-p)g(n) + p)^2}{n}\right) \nonumber
        \\ & = O\left(n\exp\left(-n^{2a-1}\right)\right) 
    } 
    by Hoeffding's inequality.
    Thus, there is a $1/\text{poly}(n)$ gap between $K_{s, x}[\lceil(1-p) \kappa(n) \rceil ]$ and $K_{s', x}[\lceil(1-p) \kappa(n) \rceil ]$ so $s$ can be distinguished from $s'$ using $\text{poly}(n)$ traces by Corollary \ref{cor:distinguish_general}.    
    
    \subsection{Proof of (\ref{eq:kdeck_degree_relation})}
    For $p < 0.5$, 
    \al{
        & 1 + c \left(\frac{ (1-p)H(1- p/(1-p)) + p\log(p/(1-p))}{1/2 - p} + 2\log\left(\frac{1}{1-p}\right) \right) \nonumber
        \\ & \leq 1 + c \left(\frac{ 1-p }{1/2 - p}\right) + 2c\log(2) \nonumber
        \\ & = 1 + c \left(\frac{ 1-p }{1/2 - p}\right) + c\log(4) \nonumber
        \\& < 4 + 12c \left(\frac{e^2}{1/2 - p}\right)    + c\log(4).
    }

    \subsection{Proof that initial $k$-subword deck algorithm from \cite{Servedio2} requires $\text{poly}(n)$ traces}
	McDiarmid's inequality states the following.  Let $X_1, ..., X_m$ are iid random variables. Suppose the function $f$ has the property that for any sets of values $x_1, x_2, ..., x_m$ and $x'_1, x'_2, ..., x'_m$ that only differ at the $i$th random variable, it follows that 
	\begin{align}
		&
		|f(x_1, x_2, ..., x_m) - f(x'_1, x'_2, ..., x'_m)| \leq b_i.
	\end{align} 
	Then, for any $t > 0$, we have that 
	\begin{align}
		\Pr(|f(X_1, X_2, ..., X_m) - \mathbb{E}[f(X_1, X_2, ..., X_m)]| \geq t) \leq 2 \exp \left(-\frac{2t^2}{\sum_{i = 1}^m b_i^2}\right). 
	\end{align}
	Applying this to our problem, we set $f = \hat{N}_x$ and we take each random variable $X_i$ to be an indicator of whether a particular bit in $s$ is deleted in a particular trace.  
	
    Recall that our estimator is given by 
    \begin{align}
        \hat{N}_x = \frac{1}{(1-p)^k}\left( \hat{E}_x - \sum_{i = k+1}^n \sum_{y \in Y_i (x)} (-1)^{|y| - |x| + 1} \hat{E}_y  \binom{y}{x}' \left(\frac{p}{1-p}\right)^{i-k} \right).
	\end{align}
	Changing the indicator random variable corresponding to a single bit in the $j$th trace $\tilde{S}_j$ changes the estimator by at most
	\begin{align}
	    b & \leq  \frac{1}{T} \frac{1}{(1-p)^k}\left( \min(k, n-k+1) + \sum_{i = k+1}^n 2 \min(i, n-i+1) \max_{y \in Y_i (x)} \binom{y}{x}' \left(\frac{p}{1-p}\right)^{i-k}\right) \nonumber
	    \\ & \leq \frac{1}{T} \frac{1}{(1-p)^k}\left( k + \sum_{i = k+1}^n 2 i  \binom{i-2}{k-2} \left(\frac{p}{1-p}\right)^{i-k}\right) \nonumber
	    \\ & \leq \frac{1}{T} \frac{1}{(1-p)^k}\left( k + 2 n \max_{i \in [k+1, n]} i  \binom{i-2}{k-2} \left(\frac{p}{1-p}\right)^{i-k}\right) \nonumber
	    \\ & \leq \frac{1}{T} \frac{1}{(1-p)^k}\left( k + 2 n^2 \max_{i \in [k+1, n]}  \binom{i-2}{k-2} \left(\frac{p}{1-p}\right)^{i-k}\right).
	\end{align}
	To analyze $b$ when $k = c \log(n)$ for $c$ constant, and $p < 0.5$ we have that
    \begin{align} b & \leq \frac{1}{T} \frac{1}{(1-p)^{c\log(n)}}\left( c\log(n) + 2 n^2 \max_{i \in [c\log(n)+1, n]}  \binom{i-2}{c\log(n)-2} \left(\frac{p}{1-p}\right)^{i-c\log(n)}\right) \nonumber
    \\ & \leq \frac{1}{T} \frac{1}{(1-p)^{c\log(n)}}\left( c\log(n) + 2 n^2 \max_{i \in [c\log(n)+1, n]}  \binom{i}{c\log(n)} \left(\frac{p}{1-p}\right)^{i-c\log(n)}\right) \nonumber
    \\ & = \frac{1}{T} \frac{1}{n^{c\log(1-p)}}\left( c\log(n) + 2 n^{2 + c \log(\frac{1-p}{p})}  \max_{i \in [c\log(n)+1, n]}  \binom{i}{c\log(n)} \left(\frac{p}{1-p}\right)^{i}\right).
	\end{align}
	We have that
	\begin{align}
	    & \max_{i \in [c\log(n)+1, n]}  \binom{i}{c\log(n)} \left(\frac{p}{1-p}\right)^{i} \nonumber
	    \\ & \leq \max_{i \in [c\log(n)+1, n]} 2^{i H(c\log(n)/i)} \left(\frac{p}{1-p}\right)^{i}
	    \nonumber
	    \\ & \leq 2^{ \frac{\log(n^c)}{1- p/(1-p)} H(1- p/(1-p))} \left(\frac{p}{1-p}\right)^{\frac{\log(n)} {1 - p/(1-p)}}
	    \nonumber
	    \\ & = n^{\frac{c H(1- p/(1-p))}{1- p/(1-p)} + \frac{c \log(p/(1-p))}{1 - p/(1-p)}} 
	\end{align}
	The maximizer above is given by $i^* = \frac{\log(n^c)}{1- p/(1-p)}$ as proved in Lemma \ref{lem:density_map_mcdiamids}.
	This is because $x = \frac{\log(n^c)}{1- q}$ is the only zero of
    \begin{align}
        & \frac{d}{dx} 2^{x H(c\log(n)/x)} q^{x} \nonumber \\ & = q^x n^{c\log(1 - \log(n^c)/x) - c\log(\log(n^c)/x)} \left(1 - \frac{\log(n^c)}{x}\right)^{-x} (\log(q) - \log(1 - \log(n^c)/x))
    \end{align}	
	where $q = p/(1-p)$,
	the function is a differentiable for $x \in (c\log(n)+1, n),$
	and the second derivative of the function at $x = \frac{\log(n^c)}{1- q}$ is given by 
    \begin{align}
        -\frac{(q-1)^2 n^{c\log(q) - c \log(1-q)} }{q\log(n^c)}
    \end{align}
    which is negative.
    We therefore have that 
        \begin{align} b &
        \leq  \frac{1}{T} \frac{1}{n^{c\log(1-p)}}\left( c\log(n) + 2 n^{2 + c \log((1-p)/p)}  n^{\frac{c H(1- p/(1-p))}{1- p/(1-p)} + \frac{c\log(p/(1-p))}{1 - p/(1-p)}} \right) \nonumber
        \\ & = \frac{1}{T} \frac{1}{n^{c\log(1-p)}}\left( c\log(n) + 2 n^{2 + c \log((1-p)/p) + \frac{c H(1- p/(1-p))}{1- p/(1-p)} + \frac{c\log(p/(1-p))}{1 - p/(1-p)}} \right)
	\end{align}    
    when $k = c\log(n)$ for $c$ constant and $p < 0.5$.
    Let
    \begin{align}
        \omega_c(p) = 2 + c \log((1-p)/p) + \frac{c H(1- p/(1-p))}{1- p/(1-p)} + \frac{c\log(p/(1-p))}{1 - p/(1-p)}
    \end{align}
	Plugging this into McDiarmid's inequality, and setting $\epsilon = 1/2$,  we have
	\begin{align}
		\Pr \left(\hat{N}_x - N_x \geq \frac{1}{2} \right) & \leq \exp \left(-\frac{2\left( \frac{1}{2}\right)^2}{nT  \left(\frac{1}{T} \frac{1}{n^{c\log(1-p)}}\left( c\log(n) + 2 n^{\omega_c(p) }  \right) \right)^2}\right) \nonumber
		\\ & = \exp \left(-\frac{T}{2n \left(\frac{1}{n^{c\log(1-p)}}\left( c\log(n) + 2 n^{\omega_c(p) }  \right)\right)^2}\right)
    \end{align}
    Setting $\delta = \Pr \left(\hat{N}_x - N_x \geq \frac{1}{2} \right)$, we have
    \begin{align}           \log(1/\delta) 2n \left(\frac{1}{n^{c\log(1-p)}}\left( c\log(n) + 2 n^{\omega_c(p) }  \right)\right)^2
    \end{align}
    traces suffice for recovering $N_x$ with probability at least $1 - \delta$.

	There are $2^{c\log(n)}$ strings $x$ in total to test, so if we want to upper bound the probability of making making a mistake in estimating any $x$ using the number of traces above, we can use the union bound. Setting $\delta = \frac{1}{n2^{c\log(n)}}$, we have
    \begin{align}
        & \log(n2^{c\log(n)})2n \left(\frac{1}{n^{c\log(1-p)}}\left( c\log(n) + 2 n^{\omega_c(p)}    \right)\right)^2 \nonumber
        \\ & = \log(e^{log(n)+ c\log(n) \log(2)}) 2n \left(\frac{1}{n^{c\log(1-p)}}\left( c\log(n) + 2 n^{\omega_c(p) } \right)\right)^2 \nonumber
        \\ & = 2n (log(n)+ c\log(n) \log(2))  \left(\frac{1}{n^{c\log(1-p)}}\left( c\log(n) + 2 n^{\omega_c(p) } \right)\right)^2 \nonumber
        \\ & = O(\text{poly}(n))
    \end{align}	
    traces suffice. 
    
    Using the union bound on the probability of error over all $2^{c\log(n)}$ kmers, we get a probability of error upper bounded by
	\begin{align}
		2^{c\log(n)} \frac{1}{n2^{c\log(n)}} = \frac{1}{n}.
	\end{align}
	Thus for $p < 0.5$, we reconstruct the  the $c\log(n)$-spectrum of $s$ with probability greater than $1 - \frac{1}{n}$ using $O(\text{poly}(n))$ traces.  	

\subsection{Expansion of (\ref{eq:recursive_formula})}
We carry out one recursive step of the expansion of (\ref{eq:recursive_formula}) in order to help illuminate the argument:
\al{
    K_{s, x}[i] & = \frac{1}{(1-p)^k} 
    \left( P_{s, x}[i] - \sum_{\ell = k+1}^n \sum_{y \in Y_\ell(x)} (1-p)^\ell \binom{y}{x}'  K_{s, y}[i] \left(\frac{p}{1-p}\right)^{\ell - k} \right) \nonumber
    \\ & = \frac{1}{(1-p)^k} 
    \Bigg( P_{s, x}[i] - \sum_{\ell = k+1}^n \sum_{y \in Y_\ell(x)} (1-p)^\ell \binom{y}{x}'   \nonumber
    \\ & \quad
    \cdot \frac{1}{(1-p)^\ell} 
    \left( P_{s, y}[i] - \sum_{j = \ell+1}^n \sum_{z \in Y_j(y)} (1-p)^j \binom{z}{y}'  K_{s, z}[i] \left(\frac{p}{1-p}\right)^{j - \ell } \right) \left(\frac{p}{1-p}\right)^{\ell - k} \Bigg)
    \nonumber
    \\ & =   \frac{1}{(1-p)^k} 
    \Bigg( P_{s, x}[i] - \sum_{\ell = k+1}^n \sum_{y \in Y_\ell(x)}  \binom{y}{x}'  P_{s, y}[i] \left(\frac{p}{1-p}\right)^{\ell - k}
    \nonumber
    \\ & \quad + 
     \sum_{\ell = k+1}^n \sum_{y \in Y_\ell(x)}  \binom{y}{x}' 
    \left( \sum_{j = \ell+1}^n \sum_{z \in Y_j(y)} (1-p)^j \binom{z}{y}'  K_{s, z}[i] \left(\frac{p}{1-p}\right)^{j - k } \right) \Bigg).
    \nonumber
}
Observe that as we expand (\ref{eq:recursive_formula}) 
one step at a time to eventually obtain (\ref{eq:formula_with_a}), every time we obtain a new term involving $P_{s, y}[i]$ in the expansion  with coefficient $c_y \left(\frac{p}{1-p}\right)^{|y| - k}$, in the next step of the expansion we obtain a term involving $P_{s, z}[i]$ with coefficient $-c_y \binom{z}{y}' \left(\frac{p}{1-p}\right)^{|z| - k}$ for every $z \in \cup_{\ell = |y| + 1}^n Y_\ell(y)$.

\end{document}